\documentclass[twocolumn,letterpaper,aps,prc,longbibliography,superscriptaddress,nofootinbib,floatfix]{revtex4-1}

\usepackage{xspace}	
\usepackage{amsmath}
\usepackage{graphicx}	

\begin{document}

\title{Measurement of jet-medium interactions via direct photon-hadron 
correlations in Au$+$Au and $d$$+$Au collisions at $\sqrt{s_{_{NN}}}=200$ GeV}

\newcommand{\abilene}{Abilene Christian University, Abilene, Texas 79699, USA}
\newcommand{\augie}{Department of Physics, Augustana University, Sioux Falls, South Dakota 57197, USA}
\newcommand{\banaras}{Department of Physics, Banaras Hindu University, Varanasi 221005, India}
\newcommand{\barc}{Bhabha Atomic Research Centre, Bombay 400 085, India}
\newcommand{\baruch}{Baruch College, City University of New York, New York, New York, 10010 USA}
\newcommand{\bnlcoll}{Collider-Accelerator Department, Brookhaven National Laboratory, Upton, New York 11973-5000, USA}
\newcommand{\bnlphys}{Physics Department, Brookhaven National Laboratory, Upton, New York 11972-5000, USA}
\newcommand{\caucr}{University of California-Riverside, Riverside, California 92521, USA}
\newcommand{\charlesczech}{Charles University, Ovocn\'{y} trh 5, Praha 1, 116 36, Prague, Czech Republic}
\newcommand{\ciae}{Science and Technology on Nuclear Data Laboratory, China Institute of Atomic Energy, Beijing 102413, People's Republic of China}
\newcommand{\cns}{Center for Nuclear Study, Graduate School of Science, University of Tokyo, 7-3-1 Hongo, Bunkyo, Tokyo 113-0033, Japan}
\newcommand{\colorado}{University of Colorado, Boulder, Colorado 80309, USA}
\newcommand{\columbia}{Columbia University, New York, New York 10027 and Nevis Laboratories, Irvington, New York 10533, USA}
\newcommand{\czechtech}{Czech Technical University, Zikova 4, 166 36 Prague 6, Czech Republic}
\newcommand{\dapnia}{Dapnia, CEA Saclay, F-91191, Gif-sur-Yvette, France}
\newcommand{\debrecen}{Debrecen University, H-4010 Debrecen, Egyetem t{\'e}r 1, Hungary}
\newcommand{\elte}{ELTE, E{\"o}tv{\"o}s Lor{\'a}nd University, H-1117 Budapest, P{\'a}zm{\'a}ny P.~s.~1/A, Hungary}
\newcommand{\eszterhazy}{Eszterh\'azy K\'aroly University, K\'aroly R\'obert Campus, H-3200 Gy\"ongy\"os, M\'atrai \'ut 36, Hungary}
\newcommand{\ewha}{Ewha Womans University, Seoul 120-750, Korea}
\newcommand{\famu}{Florida A\&M University, Tallahassee, FL 32307, USA}
\newcommand{\fit}{Florida Institute of Technology, Melbourne, Florida 32901, USA}
\newcommand{\fsu}{Florida State University, Tallahassee, Florida 32306, USA}
\newcommand{\gsu}{Georgia State University, Atlanta, Georgia 30303, USA}
\newcommand{\hanyang}{Hanyang University, Seoul 133-792, Korea}
\newcommand{\hiroshima}{Hiroshima University, Kagamiyama, Higashi-Hiroshima 739-8526, Japan}
\newcommand{\ihepprot}{IHEP Protvino, State Research Center of Russian Federation, Institute for High Energy Physics, Protvino, 142281, Russia}
\newcommand{\illuiuc}{University of Illinois at Urbana-Champaign, Urbana, Illinois 61801, USA}
\newcommand{\inrras}{Institute for Nuclear Research of the Russian Academy of Sciences, prospekt 60-letiya Oktyabrya 7a, Moscow 117312, Russia}
\newcommand{\instpasczech}{Institute of Physics, Academy of Sciences of the Czech Republic, Na Slovance 2, 182 21 Prague 8, Czech Republic}
\newcommand{\isu}{Iowa State University, Ames, Iowa 50011, USA}
\newcommand{\jaea}{Advanced Science Research Center, Japan Atomic Energy Agency, 2-4 Shirakata Shirane, Tokai-mura, Naka-gun, Ibaraki-ken 319-1195, Japan}
\newcommand{\jeonbuk}{Jeonbuk National University, Jeonju, 54896, Korea}
\newcommand{\jinrdubna}{Joint Institute for Nuclear Research, 141980 Dubna, Moscow Region, Russia}
\newcommand{\jyvaskyla}{Helsinki Institute of Physics and University of Jyv{\"a}skyl{\"a}, P.O.Box 35, FI-40014 Jyv{\"a}skyl{\"a}, Finland}
\newcommand{\kek}{KEK, High Energy Accelerator Research Organization, Tsukuba, Ibaraki 305-0801, Japan}
\newcommand{\korea}{Korea University, Seoul, 02841}
\newcommand{\kurchatov}{National Research Center ``Kurchatov Institute", Moscow, 123098 Russia}
\newcommand{\kyoto}{Kyoto University, Kyoto 606-8502, Japan}
\newcommand{\labllr}{Laboratoire Leprince-Ringuet, Ecole Polytechnique, CNRS-IN2P3, Route de Saclay, F-91128, Palaiseau, France}
\newcommand{\lahorelums}{Physics Department, Lahore University of Management Sciences, Lahore 54792, Pakistan}
\newcommand{\lawllnl}{Lawrence Livermore National Laboratory, Livermore, California 94550, USA}
\newcommand{\losalamos}{Los Alamos National Laboratory, Los Alamos, New Mexico 87545, USA}
\newcommand{\lpc}{LPC, Universit{\'e} Blaise Pascal, CNRS-IN2P3, Clermont-Fd, 63177 Aubiere Cedex, France}
\newcommand{\lund}{Department of Physics, Lund University, Box 118, SE-221 00 Lund, Sweden}
\newcommand{\maryland}{University of Maryland, College Park, Maryland 20742, USA}
\newcommand{\mass}{Department of Physics, University of Massachusetts, Amherst, Massachusetts 01003-9337, USA}
\newcommand{\michigan}{Department of Physics, University of Michigan, Ann Arbor, Michigan 48109-1040, USA}
\newcommand{\muenster}{Institut f\"ur Kernphysik, University of M\"unster, D-48149 M\"unster, Germany}
\newcommand{\muhlenberg}{Muhlenberg College, Allentown, Pennsylvania 18104-5586, USA}
\newcommand{\myongji}{Myongji University, Yongin, Kyonggido 449-728, Korea}
\newcommand{\nagasaki}{Nagasaki Institute of Applied Science, Nagasaki-shi, Nagasaki 851-0193, Japan}
\newcommand{\nara}{Nara Women's University, Kita-uoya Nishi-machi Nara 630-8506, Japan}
\newcommand{\natmephi}{National Research Nuclear University, MEPhI, Moscow Engineering Physics Institute, Moscow, 115409, Russia}
\newcommand{\newmex}{University of New Mexico, Albuquerque, New Mexico 87131, USA}
\newcommand{\nmsu}{New Mexico State University, Las Cruces, New Mexico 88003, USA}
\newcommand{\northcg}{Physics and Astronomy Department, University of North Carolina at Greensboro, Greensboro, North Carolina 27412, USA}
\newcommand{\ohio}{Department of Physics and Astronomy, Ohio University, Athens, Ohio 45701, USA}
\newcommand{\ornl}{Oak Ridge National Laboratory, Oak Ridge, Tennessee 37831, USA}
\newcommand{\orsay}{IPN-Orsay, Univ.~Paris-Sud, CNRS/IN2P3, Universit\'e Paris-Saclay, BP1, F-91406, Orsay, France}
\newcommand{\peking}{Peking University, Beijing 100871, People's Republic of China}
\newcommand{\pnpi}{PNPI, Petersburg Nuclear Physics Institute, Gatchina, Leningrad region, 188300, Russia}
\newcommand{\pusan}{Pusan National University, Pusan 46241, Korea}
\newcommand{\riken}{RIKEN Nishina Center for Accelerator-Based Science, Wako, Saitama 351-0198, Japan}
\newcommand{\rikjrbrc}{RIKEN BNL Research Center, Brookhaven National Laboratory, Upton, New York 11973-5000, USA}
\newcommand{\rikkyo}{Physics Department, Rikkyo University, 3-34-1 Nishi-Ikebukuro, Toshima, Tokyo 171-8501, Japan}
\newcommand{\saispbstu}{Saint Petersburg State Polytechnic University, St.~Petersburg, 195251 Russia}
\newcommand{\saopaulo}{Universidade de S{\~a}o Paulo, Instituto de F\'{\i}sica, Caixa Postal 66318, S{\~a}o Paulo CEP05315-970, Brazil}
\newcommand{\seoulnat}{Department of Physics and Astronomy, Seoul National University, Seoul 151-742, Korea}
\newcommand{\stonybrkc}{Chemistry Department, Stony Brook University, SUNY, Stony Brook, New York 11794-3400, USA}
\newcommand{\stonycrkp}{Department of Physics and Astronomy, Stony Brook University, SUNY, Stony Brook, New York 11794-3800, USA}
\newcommand{\sungskku}{Sungkyunkwan University, Suwon, 440-746, Korea}
\newcommand{\tenn}{University of Tennessee, Knoxville, Tennessee 37996, USA}
\newcommand{\titech}{Department of Physics, Tokyo Institute of Technology, Oh-okayama, Meguro, Tokyo 152-8551, Japan}
\newcommand{\tsukuba}{Tomonaga Center for the History of the Universe, University of Tsukuba, Tsukuba, Ibaraki 305, Japan}
\newcommand{\vandy}{Vanderbilt University, Nashville, Tennessee 37235, USA}
\newcommand{\waseda}{Waseda University, Advanced Research Institute for Science and Engineering, 17  Kikui-cho, Shinjuku-ku, Tokyo 162-0044, Japan}
\newcommand{\weizmann}{Weizmann Institute, Rehovot 76100, Israel}
\newcommand{\wigner}{Institute for Particle and Nuclear Physics, Wigner Research Centre for Physics, Hungarian Academy of Sciences (Wigner RCP, RMKI) H-1525 Budapest 114, POBox 49, Budapest, Hungary}
\newcommand{\yonsei}{Yonsei University, IPAP, Seoul 120-749, Korea}
\newcommand{\zagreb}{Department of Physics, Faculty of Science, University of Zagreb, Bijeni\v{c}ka c.~32 HR-10002 Zagreb, Croatia}
\affiliation{\abilene}
\affiliation{\augie}
\affiliation{\banaras}
\affiliation{\barc}
\affiliation{\baruch}
\affiliation{\bnlcoll}
\affiliation{\bnlphys}
\affiliation{\caucr}
\affiliation{\charlesczech}
\affiliation{\ciae}
\affiliation{\cns}
\affiliation{\colorado}
\affiliation{\columbia}
\affiliation{\czechtech}
\affiliation{\dapnia}
\affiliation{\debrecen}
\affiliation{\elte}
\affiliation{\eszterhazy}
\affiliation{\ewha}
\affiliation{\famu}
\affiliation{\fit}
\affiliation{\fsu}
\affiliation{\gsu}
\affiliation{\hanyang}
\affiliation{\hiroshima}
\affiliation{\ihepprot}
\affiliation{\illuiuc}
\affiliation{\inrras}
\affiliation{\instpasczech}
\affiliation{\isu}
\affiliation{\jaea}
\affiliation{\jeonbuk}
\affiliation{\jinrdubna}
\affiliation{\jyvaskyla}
\affiliation{\kek}
\affiliation{\korea}
\affiliation{\kurchatov}
\affiliation{\kyoto}
\affiliation{\labllr}
\affiliation{\lahorelums}
\affiliation{\lawllnl}
\affiliation{\losalamos}
\affiliation{\lpc}
\affiliation{\lund}
\affiliation{\maryland}
\affiliation{\mass}
\affiliation{\michigan}
\affiliation{\muenster}
\affiliation{\muhlenberg}
\affiliation{\myongji}
\affiliation{\nagasaki}
\affiliation{\nara}
\affiliation{\natmephi}
\affiliation{\newmex}
\affiliation{\nmsu}
\affiliation{\northcg}
\affiliation{\ohio}
\affiliation{\ornl}
\affiliation{\orsay}
\affiliation{\peking}
\affiliation{\pnpi}
\affiliation{\pusan}
\affiliation{\riken}
\affiliation{\rikjrbrc}
\affiliation{\rikkyo}
\affiliation{\saispbstu}
\affiliation{\saopaulo}
\affiliation{\seoulnat}
\affiliation{\stonybrkc}
\affiliation{\stonycrkp}
\affiliation{\sungskku}
\affiliation{\tenn}
\affiliation{\titech}
\affiliation{\tsukuba}
\affiliation{\vandy}
\affiliation{\waseda}
\affiliation{\weizmann}
\affiliation{\wigner}
\affiliation{\yonsei}
\affiliation{\zagreb}
\author{U.~Acharya} \affiliation{\gsu} 
\author{A.~Adare} \affiliation{\colorado} 
\author{S.~Afanasiev} \affiliation{\jinrdubna} 
\author{C.~Aidala} \affiliation{\losalamos} \affiliation{\mass} \affiliation{\michigan} 
\author{N.N.~Ajitanand} \altaffiliation{Deceased} \affiliation{\stonybrkc} 
\author{Y.~Akiba} \email[PHENIX Spokesperson: ]{akiba@rcf.rhic.bnl.gov} \affiliation{\riken} \affiliation{\rikjrbrc} 
\author{R.~Akimoto} \affiliation{\cns} 
\author{H.~Al-Bataineh} \affiliation{\nmsu} 
\author{J.~Alexander} \affiliation{\stonybrkc} 
\author{H.~Al-Ta'ani} \affiliation{\nmsu} 
% ...
\collaboration{PHENIX Collaboration} \noaffiliation

\date{\today}


\begin{abstract}


We present direct photon-hadron correlations in 200 GeV/A Au$+$Au, 
$d$$+$Au and $p$$+$$p$ collisions, for direct photon $p_T$ from 5--12 
GeV/$c$, collected by the PHENIX Collaboration in the years from 2006 to 
2011. We observe no significant modification of jet fragmentation in 
$d$$+$Au collisions, indicating that cold nuclear matter effects are 
small or absent. Hadrons carrying a large fraction of the quark's 
momentum are suppressed in Au$+$Au compared to $p$$+$$p$ and $d$$+$Au. 
As the momentum fraction decreases, the yield of hadrons in Au$+$Au 
increases to an excess over the yield in $p$$+$$p$ collisions. The 
excess is at large angles and at low hadron $p_T$ and is most pronounced 
for hadrons associated with lower momentum direct photons. Comparison to 
theoretical calculations suggests that the hadron excess arises from 
medium response to energy deposited by jets.

\end{abstract}

\maketitle

\section{Introduction}

Collisions of heavy nuclei at the Relativistic Heavy Ion Collider (RHIC) 
produce matter that is sufficiently hot and dense to form a plasma of 
quarks and gluons~\cite{phenix_white}. Bound hadronic states cannot 
exist in a quark gluon plasma, as the temperatures far exceed the 
transition temperature calculated by lattice quantum chromodynamics 
(QCD)~\cite{hotqcd}. Experimental measurements and theoretical analyses 
have shown that this plasma exhibits remarkable properties, including 
opacity to traversing quarks and 
gluons~\cite{PHENIXjetquench,STARawayjet}. However, the exact mechanism 
for energy loss by these partons in quark gluon plasma and the transport 
of the deposited energy within the plasma is not yet understood.

Experimental probes to address these questions include high momentum 
hadrons, reconstructed jets, and correlations among particles arising 
from hard partonic scatterings~\cite{phenix_white} occurring in the 
initial stages of the collision. Direct photons are produced dominantly 
via the QCD analog of Compton scattering, q + g $\rightarrow$ q + 
$\gamma$, at leading order, and do not interact via the strong force as 
they traverse the plasma. In the limit of negligible initial partonic 
transverse momentum, the final state quark and photon are emitted 
back-to-back in azimuth with the photon balancing the transverse 
momentum of the jet arising from the quark. Consequently, measuring the 
correlation of high momentum direct photons with opposing hadrons allows 
investigation of quark gluon plasma effects upon transiting quarks and 
their fragmentation into hadrons.

Correlations of direct photons with hadrons and jets have been measured 
by the PHENIX~\cite{ppg090,ppg113} and STAR~\cite{star_jet_corr} 
Collaborations at RHIC, and by the CMS and ATLAS collaborations at the 
Large Hadron 
Collider~\cite{CMSgammajet,ATLASgammajetFrag,ATLASxjgam,ATLASxjz,Sirunyan:2018ncy,Sirunyan:2018qec,Sirunyan:2017qhf}. Using 
the photon energy to tag the initial energy of the quark showed that 
quarks lose a substantial amount of energy while traversing the 
plasma~\cite{ppg095,ppg113}. The photon tag also allows construction of 
the quark fragmentation function $D(z)$, where 
\mbox{$z=p^{\rm hadron}/p^{\rm parton}$}. Here, $z$ represents the fraction of 
the quark's original longitudinal momentum carried by the hadrons. In 
photon-hadron ($\gamma$-h) correlations, $z$ can be approximated by 
$z_T$ = $p_T^{\sc hadron}/p_T^{\gamma}$. Comparison of $\gamma$-h 
correlations in heavy ion collisions to those in $p$$+$$p$ collisions 
quantifies the plasma's impact on parton fragmentation. $\gamma$-h 
correlations in p+A or $d$$+$$A$ collisions will reflect any cold 
nuclear matter modification of jet fragmentation. The CMS collaboration 
also studied jets correlated to neutral $Z$ bosons~\cite{Sirunyan:2017jic}.

At RHIC, the fragmentation function is substantially modified in central 
Au$+$Au collisions~\cite{ppg113,star_jet_hadron}. High $z$ fragments are 
suppressed, as expected from energy loss. Low $z$ fragments are enhanced 
at large angles with respect to the jet core, i.e. with respect to the 
original quark direction. CMS and ATLAS have measured jet fragmentation 
functions using reconstructed jets to tag the parton energy. These 
studies, conducted with jet energies of $\approx$ 100 GeV, show 
enhancement of low $p_T$ (i.e. low $z$) jet fragments in central Pb$+$Pb 
collisions~\cite{cmsFF,atlasFF}. In addition, CMS has shown that the 
energy lost by the quark is approximately balanced by hadrons with 
approximately 2 GeV $p_T$~\cite{cmsjetfragbalance} in the intrajet 
region. This is in qualitative agreement with the RHIC result, even 
though the initial quark energy differs by an order of magnitude.

There has been considerable theoretical effort to describe jet-medium 
interactions. Several mechanisms for parton energy loss were compared by 
the JET Collaboration~\cite{JET}. The medium response to deposited 
energy is now under study by several 
groups~\cite{LBT,jetscape,JEWEL_phys,JEWEL_medium}. The 
deposited energy may be totally equilibrated in the plasma, but
alternatively the deposited energy may kick up a wake in the expanding 
plasma~\cite{LBT,coLBT}. Different descriptions of plasma-modified gluon 
splitting result in different fragmentation functions, and can be tested 
by comparing the predictions to direct photon-hadron 
($\gamma_{\rm dir}$-h) correlations. 

The previously published analysis of $\gamma_{\rm dir}$-h correlations 
showed an enhancement in soft particle production at large angles. However, 
due to limited statistics, it was not possible to investigate how the 
fragmentation function depends on the parton energy or the medium scale. In 
this paper, we explore this question by looking at the direct photon $p_T$ 
dependence of the fragmentation function modification.  We investigate 
whether enhancement over the fragmentation function in $p$$+$$p$ collisions 
depends on the fragment $z_T$ or on the fragment $p_T$.  That is, does it 
depend on the jet structure or does it reflect the distribution of particles 
in the medium?  We also present first results on $\gamma_{\rm dir}$-h 
correlations in d$+$Au collisions to investigate possible cold-nuclear-matter 
effects on the fragmentation function. Fragmentation function 
modification is quantified here by the nuclear modification factor, 
$I_{AA}$, which is a ratio of the fragmentation function in Au$+$Au 
collisions to that in p$+$p collisions.

\section{Dataset and Analysis}

In 2011, PHENIX collected data from Au$+$Au collisions at 
$\sqrt{s_{_{NN}}}=200$ GeV. After event selection and quality cuts, 4.4 
billion minimum-bias (MB) events were analyzed. These are combined with the 
previously reported 3.9 billion MB Au$+$Au events from 2007 and 
2.9 billion from 2010~\cite{ppg113}. The high momentum photon triggered 
$d$$+$Au data set at $\sqrt{s_{_{NN}}}$ = 200 GeV was collected in 2008, and 3 
billion events are analyzed. The $p$$+$$p$ comparison data are from 2005 and 
2006~\cite{ppg095}.

\begin{figure}[htb]
\includegraphics[width=1.0\linewidth]{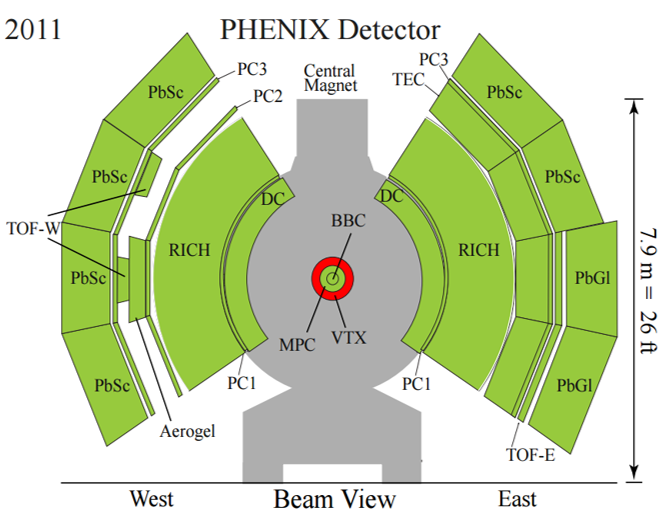}
\caption {Side view of the PHENIX central arm spectrometers in 2011.}
\label{fig:phenix}
\end{figure}

The measurements in this paper use the PHENIX central 
spectrometers~\cite{detectors}. Two particle correlations are 
constructed by pairing photons or $\pi^0$s measured in the 
electromagnetic calorimeter (EMCal)~\cite{emcal} with charged hadrons 
reconstructed in the drift chambers and pad chambers~\cite{tracking}. 
The acceptance in pseudorapidity is $|\eta|<0.35$, while each 
spectrometer arm covers 90 degrees in azimuth. Beam-beam 
counters~\cite{bbc}, located at 1.44 meters from the center of the 
interaction region, cover the pseudorapidity range from 3.0 to 3.9 and 
full azimuthal angle.  They are used to determine the collision 
centralities and vertex positions.  Figure~\ref{fig:phenix} shows the 
detector configuration in 2011.

Photons and $\pi^0$s are measured in the EMCal. There are four sectors 
of lead-scintillator (PbSc) sampling calorimeters in the west arm, while 
the east arm has two sectors of lead-scintillator and two lead-glass 
(PbGl) \v{C}erenkov calorimeters. The PbSc and PbGl calorimeters have 
energy resolutions of $\sigma_E/\sqrt{E}$ = 8.1$\%$/$\sqrt{E}$ $\oplus$ 
2.1$\%$ and 5.9$\%$/$\sqrt{E}$ $\oplus$ 0.8$\%$, respectively. Photons 
are selected via an electromagnetic shower shape cut~\cite{showershape} 
on energy clusters. The high granularity of the EMCal, $\delta\eta 
\times \delta\phi$ = 0.011 $\times$ 0.011 for PbSc and 0.008 $\times$ 
0.008 for PbGl, allows for $\pi^0$ reconstruction via the 
$\pi^0\rightarrow\gamma\gamma$ channel (invariant mass = 120--160 
MeV/c$^2$) up to $p_T$ = 15 GeV, beyond which shower merging becomes 
significant. A charged track veto is applied to remove possible hadron 
or electron contamination in the photon sample, reducing 
auto-correlations in the measurement. The EMCal system is also used to 
trigger on $d$$+$Au events with high $p_T$ photons.

Two particle correlations are constructed as a function of $\Delta\phi$, 
the azimuthal angle between photon or $\pi^0$ triggers and associated 
hadron partners. Pairs arise from jet correlations superimposed on a 
combinatorial background from the underlying event. In $p$$+$$p$ and $d$$+$Au 
collisions where the event multiplicity is low, we treat this background 
as flat in $\Delta\phi$ and subtract it, normalizing the level via the 
zero-yield-at-minimum (ZYAM) procedure~\cite{ZYAM}. In Au$+$Au collisions, 
the background has an azimuthal asymmetry quantified in the flow 
parameters $v_n$, which are used to modulate the subtracted background, 
as described in Eqn. 1. Only $v_2$ is included in the subtraction, while 
higher-order effects are included as an additional systematic 
uncertainty on the final results.

We report jet pairs as conditional (or per-trigger) yields of hadrons. 
Detector acceptance corrections are determined using mixed events with 
similar centrality and collision vertex. For Au$+$Au collisions, the 
background level $b_0$ is estimated using an absolute 
normalization~\cite{ZYAM}, determined from the uncorrelated 
single-photon and single-hadron production rates. The final invariant 
yield of associated hadrons is obtained by dividing the 
background-subtracted correlated hadron yields by the number of triggers 
$N_t$ and correcting for the associate charged hadron efficiency 
$\epsilon^a$, determined by a {\sc geant} detector simulation:

\begin{widetext}
\begin{equation}
\frac{1}{N_t}\frac{dN^{\rm pair}}{d\Delta\phi} = \frac{1}{N_t}\frac{N^{\rm pair}}{\epsilon^{a}\int\Delta\phi}\Bigg\{\frac{dN^{\rm pair}_{\rm real}/d\Delta\phi}{dN^{\rm pair}_{\rm mix}/d\Delta\phi} - b_0\big[1+2\big\langle v_2^tv_2^a \big\rangle cos\big(2\Delta\phi\big) \big] \Bigg\},
\end{equation}
\end{widetext}

\noindent where $v_2^t$ and $v_2^a$ are the elliptic flow magnitudes 
independently measured for the trigger and associated particles, 
respectively~\cite{ppg113}. These modulate the angular distribution of 
the background. Lastly, $N^{\rm pair}$ denotes the number of 
trigger-associate pairs. The subscript ``real" refers to a 
trigger-associate particle pair that came from the same event, and the 
subscript ``mix" refers to trigger-associate pairs that come from 
different events and are used to correct for correlations due to 
detector effects.

In both Au$+$Au and $d$$+$Au analyses, photons with transverse momentum of 5 
to 15 GeV/$c$ are selected as triggers.  To extract yields of 
hadrons associated with direct photons, the background from decay photon 
correlations with hadrons must be subtracted. In Au$+$Au collisions, where 
the multiplicity is high, this is achieved via a statistical subtraction 
procedure. If $N_{inc}$, $N_{dec}$, and $N_{dir}$ are the inclusive, decay, 
and direct photon yields, respectively, then $N_{dir} = N_{inc} - N_{dec}$. 
It follows that the conditional yield of hadrons $Y$ for different 
photon trigger samples is

\begin{equation}\label{eq:direct_photon_statistical_subtraction}
Y_{\rm dir} = \frac{R_{\gamma}Y_{\rm inc} - Y_{\rm dec}}{R_{\gamma}-1},
\end{equation}

\noindent where $R_{\gamma} \equiv N_{inc}/N_{dec}$ and is measured 
independently~\cite{ppg139}.




The decay photon background is estimated using measured $\pi^0$-hadron 
($\pi^0$-h) correlations and a Monte Carlo pair-by-pair mapping 
procedure. The simulation calculates the probability distribution for 
decay photon-hadron ($\gamma_{\rm dec}$-h) pairs in a certain photon $p_T$ 
range as a function of the parent $\pi^0$ $p_T$. $\gamma_{\rm dec}$-h 
correlations are constructed via a weighted sum over all individual 
$\pi^0$-hadron pairs, where the weighting factor reflects the kinematic 
probability for a $\pi^0$ at a given $p_T$ to decay into a photon in the 
selected $p_T$ range. The $\gamma_{\rm dec}$-h per-trigger yield can be 
described by the following equation:

\begin{equation}
Y_{\rm dec} = \frac{\int \rho(p_{T \pi^0} \rightarrow p_{T \gamma}) \epsilon^{-1}(p_{T \pi^0})N_{\pi^0-h}dp_{T \pi^0}}{\int \rho(p_{T \pi^0} \rightarrow p_{T \gamma}) \epsilon^{-1}(p_{T \pi^0})N_{\pi^0}dp_{T \pi^0}},
\end{equation}

\noindent where $\rho$ gives the probability that a $\pi^0$ decays to a 
photon with $p_{T\gamma}$, and $\epsilon$ is the $\pi^0$ reconstruction 
efficiency, which can be determined by scaling the raw $\pi^0$ spectra 
to a power law fit to published data~\cite{ppg080}. $N_{\pi^0-h}$ and 
$N_{\pi^0}$ are the number of $\pi^0$-h pairs and number of $\pi^0$'s, 
respectively. When reconstructing the $\pi^0$, a strict cut on the 
asymmetry of the energy of the two photons is applied to reduce the 
combinatorial background from low energy photons. The probability 
weighting function, determined from Monte Carlo simulation, takes into 
account the actual EMCal response, including energy and position 
resolution and detector acceptance.

With the $\pi^0$ to decay photon $p_T$ map, $\rho$, the inclusive photon 
sample can be separated into a meson decay component and a direct 
component. Of the meson decay component, 20$\%$ of the decay yield is from non-$\pi^0$ decays as calculated from previous PHENIX results~\cite{Adler:2006bv}.
To construct $\gamma_{\rm dec}$-h yields with trigger photon 
$p_T$ of 5--15 GeV/$c$, hadron correlations with $\pi^0$ of 4 $\leq p_T 
\leq$ 17 GeV/$c$ are utilized. The slightly wider $p_T$ range is chosen 
to account for decay kinematics, as well as $p_T$ smearing from the 
EMCal energy and position resolution. An additional cutoff correction 
accounts for the small $\gamma_{\rm dec}$-h yield in the trigger $p_T$ 
range 5--15 GeV/$c$ from $\pi^0$ with $p_T \geq$ 17 GeV/$c$. The merging 
of decay photons from high $p_T$ $\pi^0$ is not accounted for in the 
Monte Carlo mapping simulation. Instead, the efficiency to detect 
photons from a high momentum parent meson is calculated via 
{\sc geant} simulation of the full detector response. This 
loss is included in the probability function as an additional 
correction. The opening angle of photon pairs that merge is small, thus 
they are removed from the measured inclusive photon sample by the shower 
shape cut.

In $d$$+$Au collisions, where the underlying event background is much 
smaller, it is possible to improve the signal to background for direct 
photons. This is done event-by-event using a photon isolation cut and by 
removing all photons identified (tagged) as resulting from a $\pi^0$ 
decay~\cite{ppg095}. First, all photons with $p_T \geq$ 0.5 GeV/$c$ are 
paired. Those pairs with invariant mass between 120--160 MeV/c$^2$ are 
tagged as decay photons and removed from the inclusive sample. Next, an 
isolation criterion is applied to the remaining photons to further 
reduce the background of decay photons, as well as contamination from 
fragmentation photons. The isolation cut requires that the energy in a 
cone around the trigger photon be less than 10$\%$ of the photon energy 
in $p$$+$$p$ collisions. In the $d$$+$Au analysis, the cut is modified slightly to 
include the effect of the modest underlying event. The underlying event 
is evaluated separately for each $d$$+$Au centrality class, resulting in an 
isolation criterion:

\begin{equation}
\sum_{\Delta R < R_{\rm max}} E < (E_{\gamma} * 0.1+<E_{bg}>),
\end{equation}
where $E$ is the measured energy in the isolation cone, $E_\gamma$ is the 
photon energy, $\Delta R$ = $\sqrt{\Delta\phi^2 + \Delta\eta^2}$ is the 
distance between the trigger photon and other particles in the event and 
$<E_{bg}>$ is the average energy inside the cone in the underlying event. 
The cone size ($R_{\rm max}$) used in this analysis is 0.4.

To account for the $d$$+$Au underlying event, the ZYAM 
procedure~\cite{ZYAM} is applied to the angular correlation functions 
for each centrality class.  As an isolation cut distorts the near-side 
yield, the minimum point is determined within the restricted 
$\Delta\phi$ range of 0.9--1.6 rad.  The zero-point yield is determined 
by integrating in a 0.03 rad range around the minimum point. The hadron 
conditional yield reported here is corrected for the PHENIX hadron 
acceptance. The ZYAM subtracted inclusive and decay yields for each 
centrality are combined using a weighted sum based on the number of each 
type of trigger to obtain the MB yields.

Some decay photons are missed by the $\pi^0$ tagging procedure and slip 
through the isolation cut to be counted as direct photons. Such falsely 
isolated $\gamma_{\rm dec}$-h correlations are corrected via a statistical 
subtraction, similar to Eq. \ref{eq:direct_photon_statistical_subtraction}. 
If we define $N_{\rm inc-tag}^{\rm iso}$ as the yield of isolated photons 
after removing those isolated photon tagged as decay photons, $N_{\rm 
dec}^{\rm miss,iso}$ as those decay photons that are isolated but not tagged 
as decay photons, and $N_{\rm dir}^{\rm iso}$ as isolated direct photons, 
then $N_{\rm dir}^{\rm iso} = N_{\rm inc-tag}^{\rm iso} - N_{\rm dec}^{\rm 
miss,iso}$. It follows that the condition of yield of hadrons for direct, 
isolated photons is

\begin{equation}
Y_{\rm dir}^{\rm iso} 
= \frac{R_{\gamma}^{\rm eff}Y_{\rm inc-tag}^{\rm iso}-Y_{\rm dec}^{\rm miss,iso}}
{R_{\gamma}^{\rm eff}-1},
\end{equation}

\noindent where
\begin{equation}
R_{\gamma}^{\rm eff} = \frac{N^{\rm iso}_{\rm inc}}{N^{\rm iso}_{\rm dec}} 
= \frac{R_\gamma} 
{(1-\epsilon^{\rm tag}_{\rm dec})(1-\epsilon^{\rm iso}_{\rm dec})}
\frac{N_{\rm inc-tag}^{\rm iso}}{N_{\rm inc}}
\end{equation}

\noindent where $\epsilon_{\rm dec}^{\rm iso}$ is the isolation cut 
efficiency and $\epsilon_{\rm dec}^{\rm tag}$ is the tagging efficiency. 
More detail on the subtraction procedures and cuts can be found in 
references~\cite{ppg090,ppg095}.

In the Au$+$Au analysis, there are four main sources of systematic 
uncertainties. The systematic uncertainty coming from the statistical 
subtraction method is due to the statistical and systematic 
uncertainties on the value of $R_{\gamma}$. There are also uncertainties 
when extracting the jet functions due to uncertainties on the value of 
the elliptic flow modulation magnitude, $v_2$. This analysis uses 
published values and uncertainties from PHENIX~\cite{ppg113}. The 
absolute normalization method to determine the underlying event 
background level, and the determination of the decay photon $p_T$ 
mapping are also significant contributors to the overall systematic 
uncertainties. The uncertainties, along with their $p_T$ and centrality 
dependence, are propagated into the final jet functions and per-trigger 
hadron yields. The systematic uncertainty on the hadron efficiency 
determination comes in as a global scale uncertainty on the correlated 
hadron yields.

In MB $d$$+$Au collisions, $v_2$ is small. However, the systematic 
uncertainties on $\gamma$-h correlations include those arising from the 
ZYAM procedure used to determine the combinatorial background. There is 
also an uncertainty arising from the $\pi^0$ tagging and isolation cuts, 
which is included in the quoted systematic uncertainty.

\begin{figure}[htb]
\includegraphics[width=1.0\linewidth]{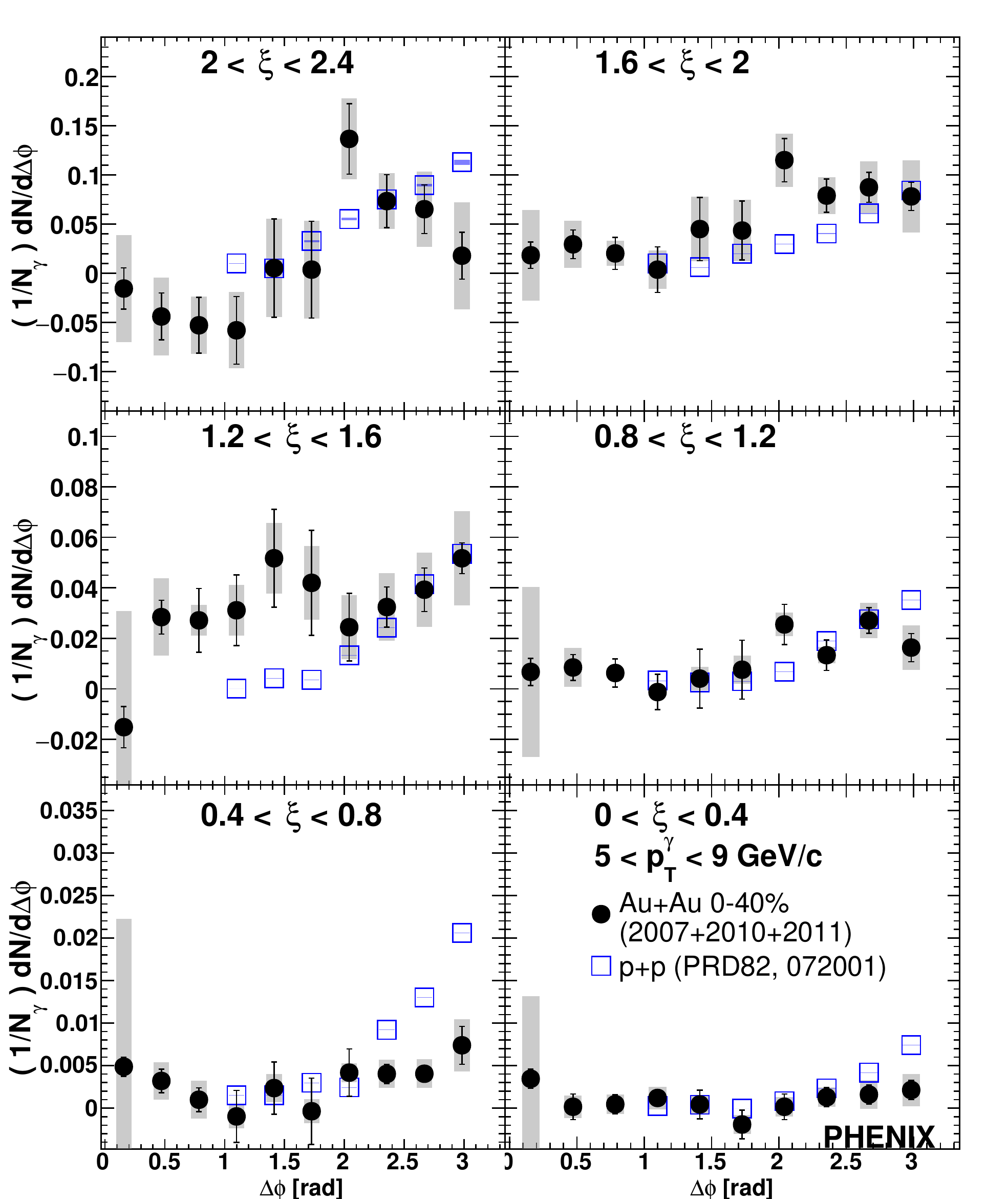}
\caption {\label{fig:cdir_5_9}
Per-trigger yield of hadrons associated to direct photons in Au$+$Au 
collisions (closed [black] circles) for direct photon $p_T$ 5--9 GeV/$c$, 
compared with $p$$+$$p$ baseline (open [blue] squares), in various $\xi$ bins.
}
\end{figure}
\begin{figure}[htb]
\vspace{0.5cm}
\includegraphics[width=0.998\linewidth]{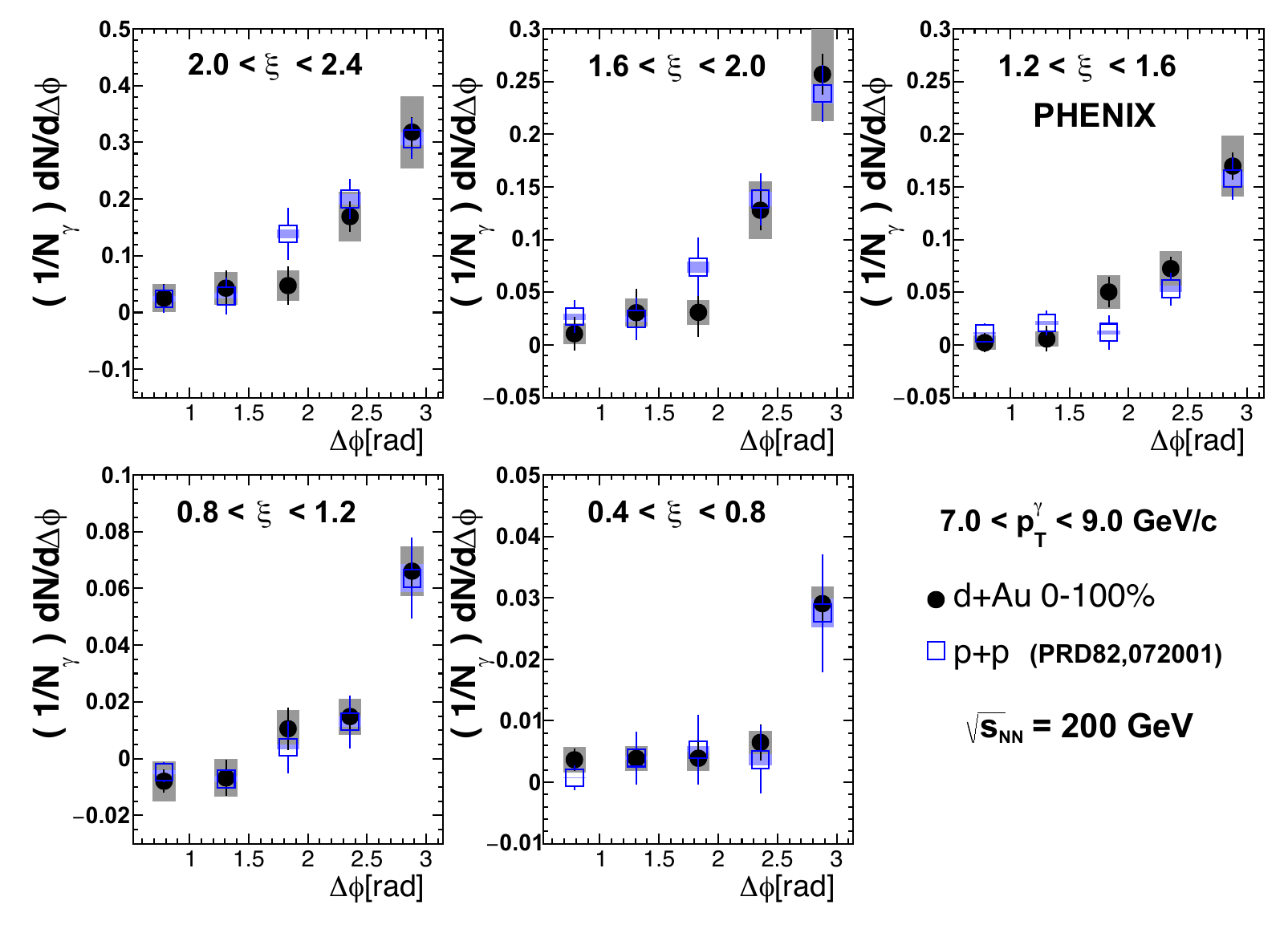}
\caption{\label{fig:cdir_dAu_1}
Per-trigger yield of hadrons associated to direct photons in $d$+$Au$ 
collisions (closed [black] circles) for direct photon $p_T$ 7--9 GeV/$c$, 
compared with $p$+$p$ baseline (open [blue] squares), in various $\xi$ bins.
}
\end{figure}

\section{Results}

In this paper, we aim to quantify the modification of the jet 
fragmentation function $D(z)$ in Au$+$Au and $d$$+$Au collisions, compared to 
the $p$$+$$p$ baseline. The jet fragmentation function describes the 
probability of an outgoing parton to yield a hadron with momentum 
fraction $z = p^{\rm hadron}/p^{\rm parton}$. Assuming that the initial-state 
$k_T$ of partons in a nucleon has a negligible effect, then $z_{T} = 
p_{T}^{\rm hadron}/p_{T}^{\gamma}$ can be used to approximate $z$.  To focus 
on the low $z_T$ region, where modification is anticipated, we use the 
variable $\xi = ln(1/z_{T})$.

Figure~\ref{fig:cdir_5_9} shows azimuthal angular distributions of 
hadrons associated with direct photons of 5 $< p_T <$ 9 GeV/$c$, in the 
0--40$\%$ most central Au$+$Au collisions, separated into bins of $\xi$. 
These distributions are a combination of the 2007, 2010 and 2011 data 
sets. The Au$+$Au results are shown as closed [black] circles, with 
shaded boxes representing systematic uncertainties on the measurement. 
The $p$$+$$p$ $\gamma_{\rm dir}$-h results are shown in open [blue] 
squares. The $p$$+$$p$ baseline measurement combines data collected in 
2005 and 2006~\cite{ppg113,ppg095}. It should be noted that the 
isolation cut in the $p$$+$$p$ analysis makes the near-side yield not 
measurable. Consequently, the $p$$+$$p$ points with $\Delta\phi<1$ are 
not shown in these distributions.

On the near side, i.e.~$\Delta\phi < \pi/2$, the Au$+$Au 
$\gamma_{\rm dir}$-h yields are consistent with zero, indicating that the 
statistical subtraction is properly carried out and next-to-leading-order 
effects are negligible. On the away-side, i.e.~$\Delta\phi>\pi/2$,
an enhancement in the Au$+$Au data compared to $p$$+$$p$ is observed in the 
higher $\xi$ bins. As noted before, this corresponds to low $z$, where 
the observed hadrons carry a small fraction of the scattered parton's 
original momentum. In the low $\xi$ bins, the Au$+$Au per-trigger yield is 
suppressed, as expected if the parton loses energy in the medium.

Fig. \ref{fig:cdir_dAu_1} shows the $\Delta\phi$ distributions of 
isolated $\gamma_{\rm dir}$-h yields in $d$$+$Au and $p$$+$$p$ 
collisions, for direct photon $p_T$ 7--9 GeV/$c$. The $d$$+$Au and 
$p$$+$$p$ results are consistent in all the measured $\xi$ bins.

\begin{figure}[htb]
\includegraphics[width=1.0\linewidth]{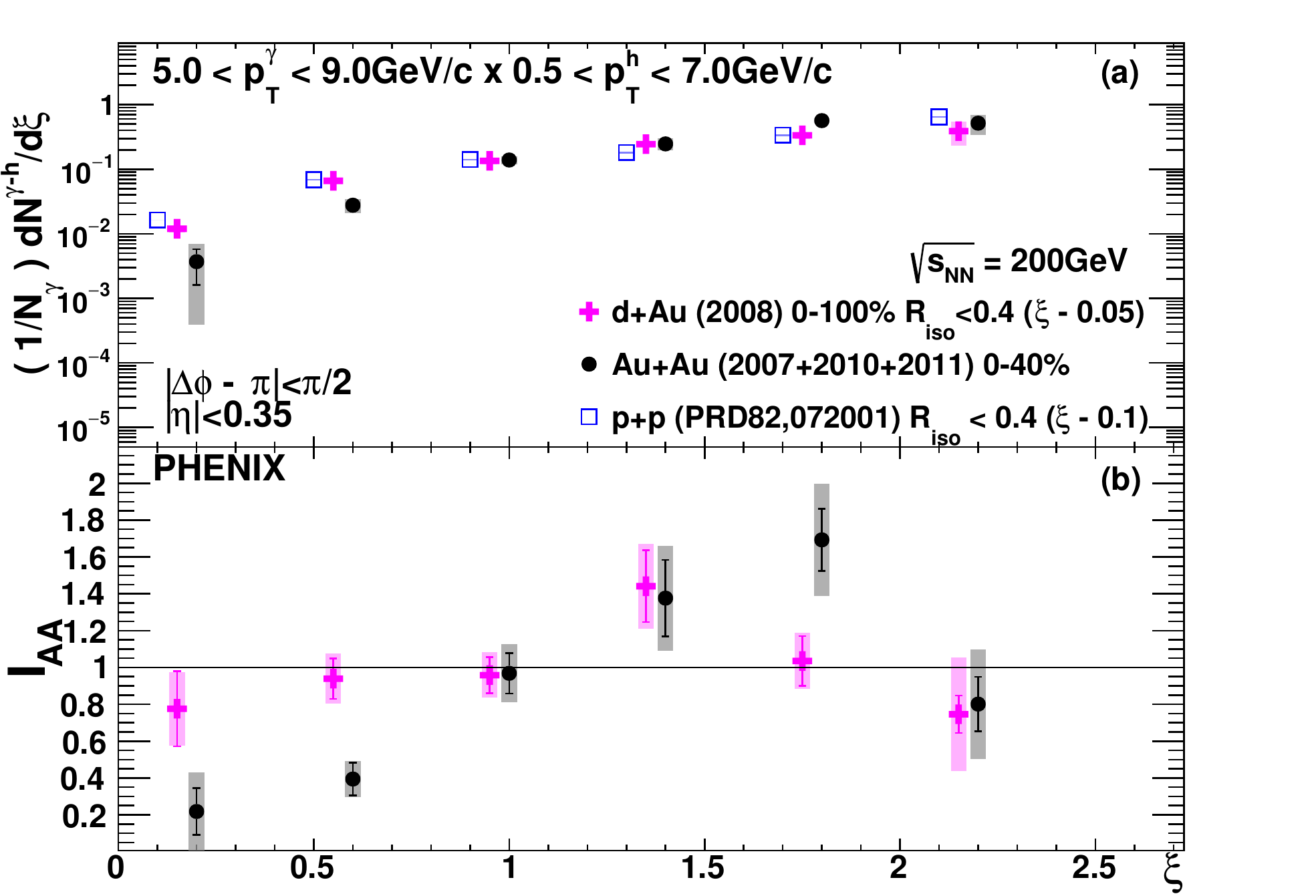}
\caption{\label{fig:iaa_ida}
(a) Integrated away-side $\gamma_{\rm dir}$-h per-trigger yields of 
Au$+$Au (closed [black] circles), $d$+$Au$ ([purple] crosses) and 
$p$+$p$ (open [blue] squares), as a function of $\xi$. The $p$+$p$ 
and $d$+$Au$ points have been shifted to the left for clear viewing, as 
indicated in the legend.  (b) $I_{AA}$ (closed [black] circles) and 
$I_{dA}$ ([purple] crosses).
}
\end{figure}

\begin{figure}[htb]
\includegraphics[width=1.0\linewidth]{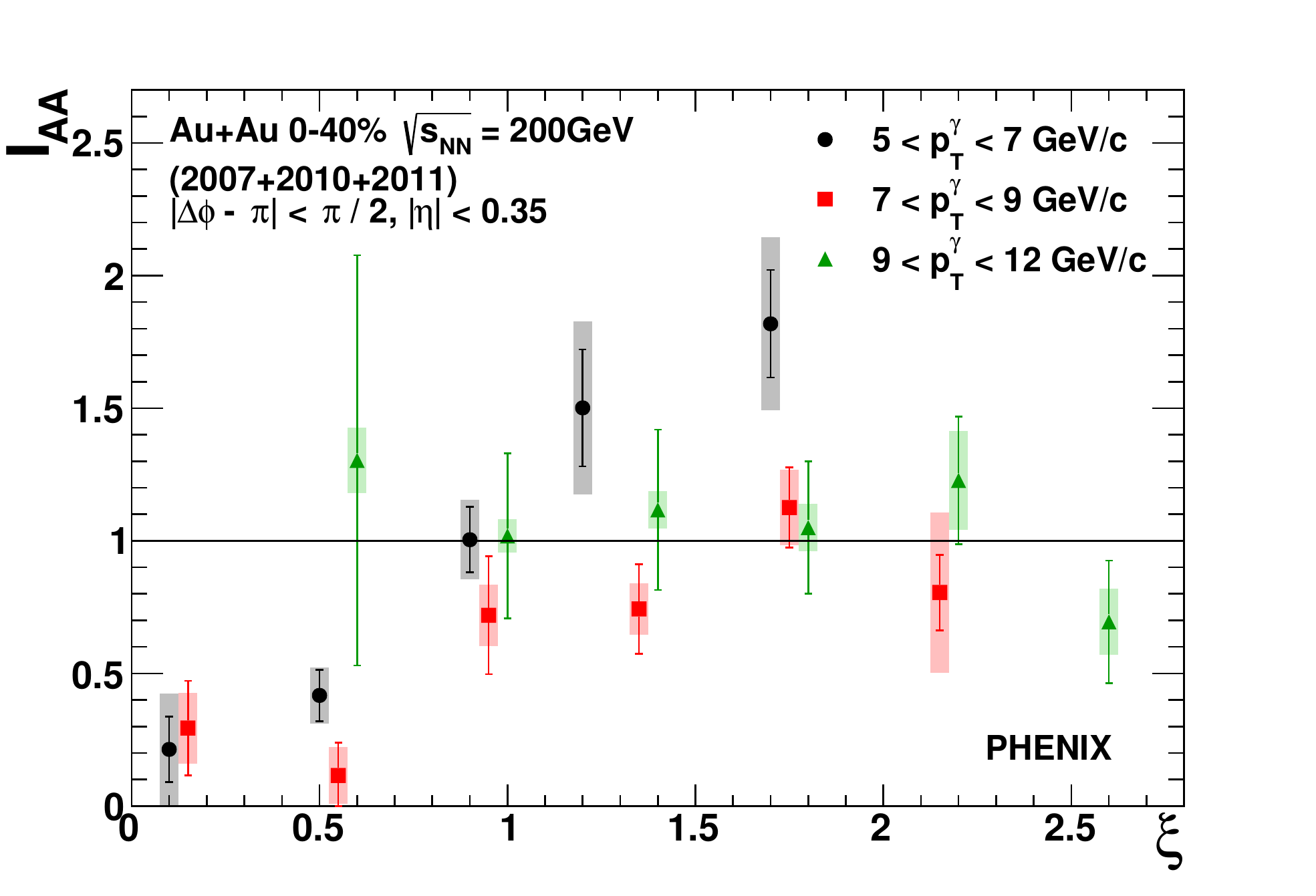}
\caption{\label{fig:iaa_3pt}
$I_{AA}$ vs $\xi$ for direct photon $p_T^{\gamma}$ of 
5--7 GeV/$c$ (closed [black] circles), 
7--9 GeV/$c$ (closed [red] squares), 
and 9--12 GeV/$c$ (closed [green] triangles).
}
\end{figure}

Figure~\ref{fig:iaa_ida}(a) shows the fragmentation 
functions for all three systems as a function of $\xi$. These are 
calculated by integrating the per-trigger yield of hadrons in the 
azimuthal angle region $|\Delta \phi - \pi| < \pi/2$~rad. Data points 
for Au$+$Au are plotted on the $\xi$ axis at the middle of each $\xi$ bin: 
0.2, 0.6, 1.0, 1.4, 1.8, 2.2. The $p$$+$$p$ and $d$$+$Au points have been shifted 
to the left in $\xi$ for viewing clarity.

As noted in the Introduction, $I_{AA} = Y_{AA}/Y_{pp}$ is a 
nuclear-modification factor, which quantifies the difference between the 
fragmentation functions in Au$+$Au and $p$$+$$p$. In the absence of any 
medium modifications, $I_{AA}$ should equal 1.

Figure~\ref{fig:iaa_ida}(b) shows $I_{AA}$ for direct photons of $5 < 
p_T^{\gamma} < 9$ GeV/$c$.  In Au$+$Au collisions, there is a clear 
suppression at low $\xi$ and enhancement at high $\xi$. The $d$$+$Au nuclear 
modification factor, $I_{dA}$, is also shown as closed [purple] crosses 
in Fig.~\ref{fig:iaa_ida}(b). $I_{dA}$ is consistent with unity across 
all $\xi$ ranges, indicating that there is no significant modification 
of the jet fragmentation function in $d$$+$Au collisions.

\begin{figure}[htb]
\includegraphics[width=1.0\linewidth]{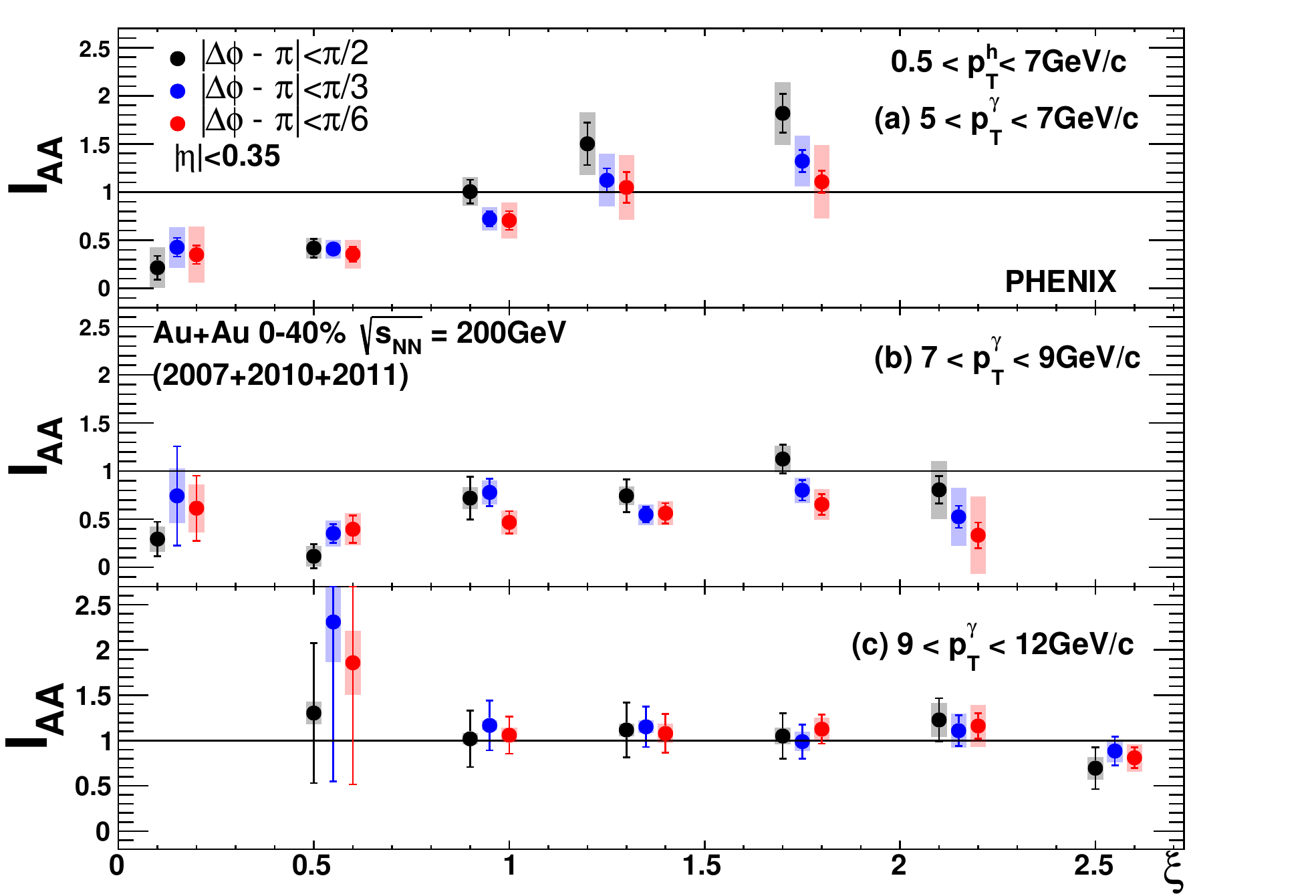}
\caption{\label{fig:iaa_3pt_3ranges}
$I_{AA}$ as a function of $\xi$ for direct photon $p_T^{\gamma}$ of 
(a) 5--7, (b) 7--9, and (c) 9--12 GeV/$c$. 
Three away-side integration ranges are chosen to calculate the 
per-trigger yield and the corresponding $I_{AA}$: 
$|\Delta\phi-\pi|<\pi/2$ (closed [black] circles), 
$|\Delta\phi-\pi|<\pi/3$ (closed [blue] squares) and 
$|\Delta\phi-\pi|<\pi/6$ (closed [red] triangles).}
\end{figure}

The statistics from the combined Au$+$Au runs allow for a differential 
measurement as a function of direct photon $p_T$ (i.e. as a function of 
the approximate jet energy). Fig. \ref{fig:iaa_3pt} shows $I_{AA}$ as a 
function of $\xi$ for three direct photon $p_T$ ranges. While the 
associated hadron yields are smaller than those in $p$$+$$p$ at low $\xi$, the 
appearance of extra particles at higher $\xi$ is observed for direct 
photons with $p_T$ of 5--7 GeV/$c$. A qualitatively similar increase of 
$I_{AA}$ with $\xi$ is visible for the 7--9 GeV/$c$ direct photon $p_T$ 
range.

To investigate where the energy deposited in the plasma goes, we study 
the dependence of $I_{AA}$ on the integration range in azimuthal opening 
angle. The hadron yields are also integrated in two narrower angular 
ranges on the away side: $|\Delta\phi-\pi| < \pi/3$ rad and 
$|\Delta\phi-\pi| < \pi/6$ rad. The resulting $I_{AA}$ values are shown 
in Fig. \ref{fig:iaa_3pt_3ranges} for all three direct photon $p_T$ 
bins. The enhancement over $p$$+$$p$ is largest for the 5--7 GeV/$c$ 
direct photon momentum range, and for the full away-side integration 
range. The suppression pattern is similar for the different integration 
regions, suggesting that the jet core is suppressed, and the enhancement 
exists at large angles. The angular distributions support the 
observation from Fig. \ref{fig:cdir_5_9}, that particle yields are 
enhanced at large angles with respect to the away-side jet axis in the 
$1.6<\xi<2.0$ bin.

\begin{figure}[htb]
\includegraphics[width=1.0\linewidth]{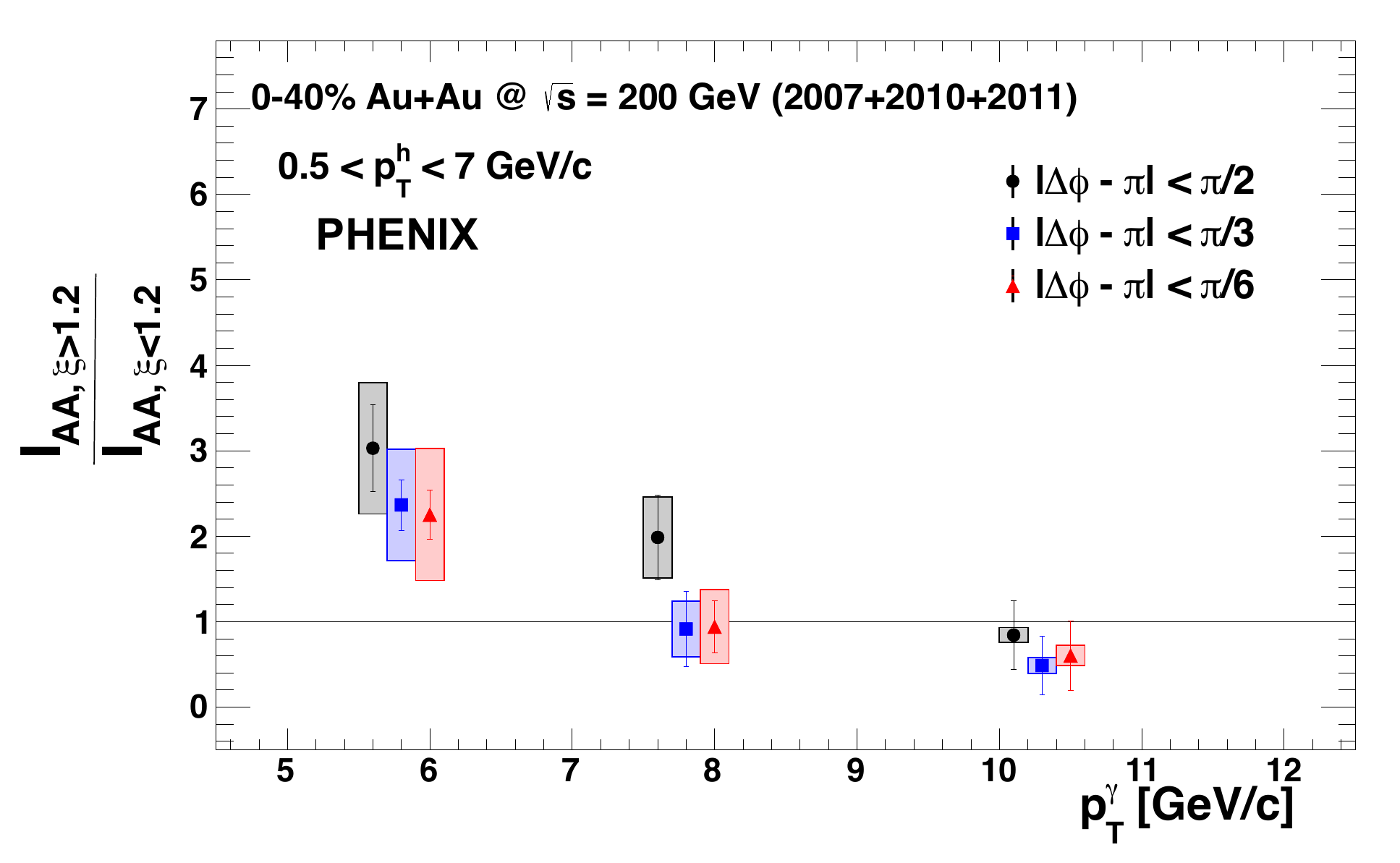}
\caption{\label{fig:iaa_hi_lo_ratio}
Ratios of $I_{AA}$ as a function of direct photon $p_T$ \\for three 
different away-side integration ranges.
}
\end{figure}

Whether or not $I_{AA}$ becomes significantly larger than unity (what 
we have been referring to as enhancement) there is a tendency for 
$I_{AA}$ to increase with increasing $\xi$. To quantify this, we 
calculate the weighted averages of $I_{AA}$ values above and below $\xi$ 
= 1.2. The ratio for each integration range is plotted in Fig. 
\ref{fig:iaa_hi_lo_ratio}, as a function of the direct photon $p_T$. The 
enhancement is largest for softer jets and for the full away-side 
integration range, implying that jets with lower energy are broadened 
more than higher energy jets.

\begin{figure}[h]
\includegraphics[width=1.0\linewidth]{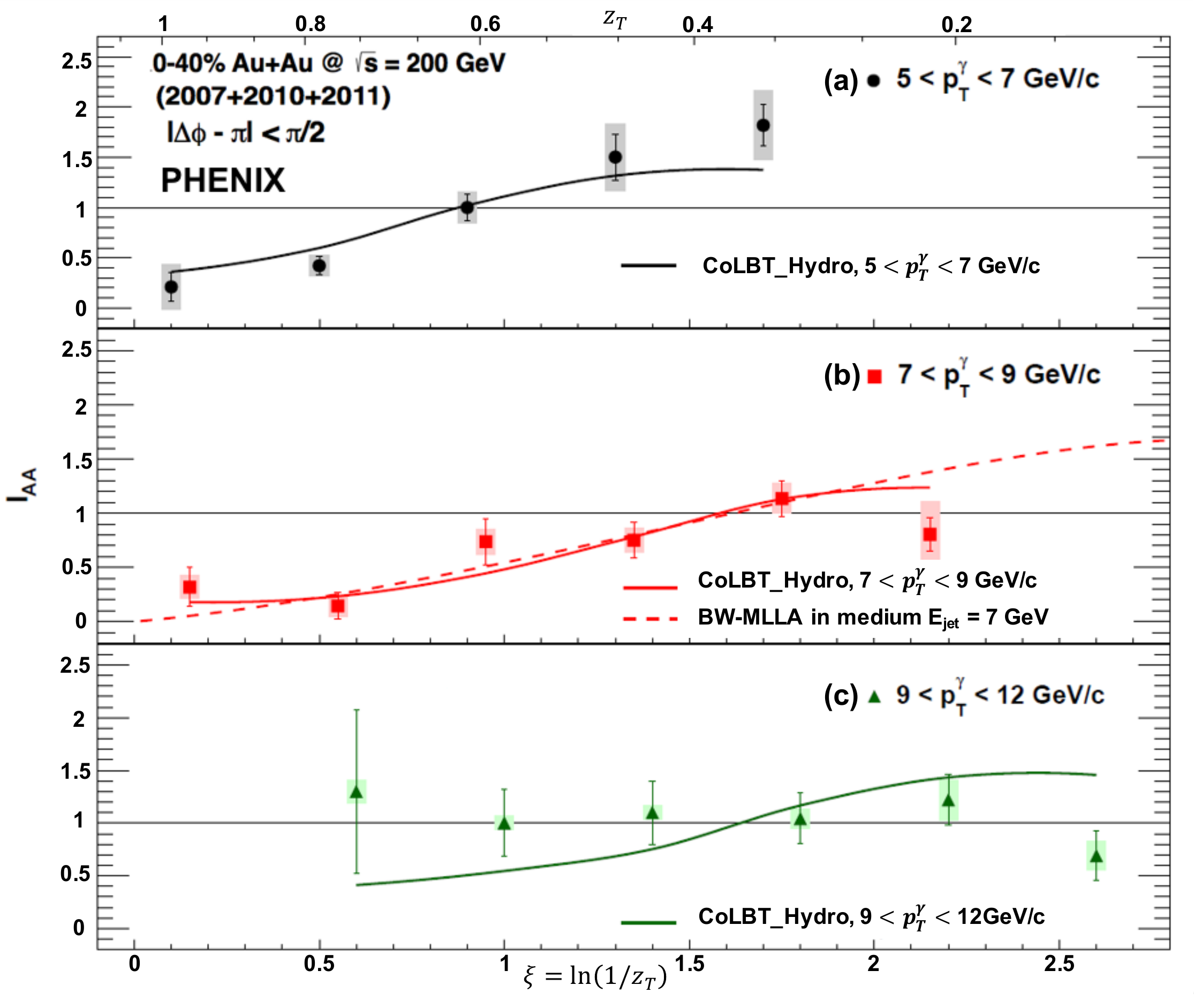}
\caption{\label{fig:iaa_theory}
Measured $I_{AA}$ for direct photon $p_T$ of (a) 5--7, (b) 7--9, and 
(c) 9--12 GeV/$c$, as a function of $\xi$, are compared with 
theoretical model calculations.
}
\end{figure}

\section {Discussion}

To determine whether $I_{dA}$ indicates any cold nuclear matter effects, 
the $\chi^2$ per degree of freedom values were calculated under the 
assumption of no modification and are determined to be 7.4/5, 4.0/5, 
10.0/5 for direct photon $p_T$ bins 5--7, 7--9, and 9--12 GeV/$c$, 
respectively. The result indicates that $I_{dA}$ is consistent with 
unity and therefore the jet fragmentation function is not significantly 
modified in $d$$+$Au collisions, within the current uncertainties. This 
suggests that any possible cold nuclear matter effect is small.

We next compare our Au$+$Au results to predictions from the CoLBT-hydro 
model~\cite{coLBT} in Fig. \ref{fig:iaa_theory}, which shows $I_{AA}$ as 
a function of $\xi$ for the 3 direct photon $p_T$ bins; the $z_T$ axis 
is displayed on the top.  The solid lines are from the CoLBT model 
calculated in the same kinematic ranges as the data. The model 
calculation shows the same trends with $\xi$ as the data . CoLBT has a 
kinetic description of the leading parton propagation, including a 
hydrodynamical picture for the medium evolution. In this calculation, 
both the propagating jet shower parton and the thermal parton are 
recorded, along with their further interactions with the medium. 
Consequently, the medium response to deposited energy is modeled. The 
model clearly shows that as the direct photon $p_T$ increases, the 
transition where $I_{AA}$ exceeds one occurs at increasing $\xi$. 
According to this calculation, the enhancement at large $\xi$ arises 
from jet-induced medium excitations, and that the enhancement occurs at 
low $z_T$ reflects the thermal nature of the produced soft particles.

Figure~\ref{fig:iaa_theory}(b) shows a BW-MLLA calculation (dashed 
[red] curve) in which it is assumed that the lost energy is 
redistributed, resulting in an enhanced production of soft 
particles~\cite{mlla}. The calculation for jets with energy of 7 
GeV in the medium is in relatively good agreement with the measured 
results.  The model comparisons suggest that the enhancement of soft 
hadrons associated with the away-side jet should scale with the $p_T$ of 
the hadrons. A modified fragmentation function could be expected to 
produce a change at fixed $z_T$. This is not consistent with either the 
data or the CoLBT model.

\section{Summary}

We have presented direct photon-hadron correlations in 
$\sqrt{s_{_{NN}}}=200$~GeV Au$+$Au, $d$$+$Au and $p$$+$$p$ collisions, for photon 
$p_T$ from 5--12 GeV/$c$. As the dominant source of correlations is QCD 
Compton scattering, we use the photon energy as a proxy for the opposing 
quark's energy to study the jet fragmentation function. Combining data 
sets from three years of data taking at RHIC allows study of the 
conditional hadron yields opposite to the direct photons as a function 
of $z_T$ and the photon $p_T$. This is the first time such a 
differential study of direct photon-hadron correlations has been 
performed at RHIC.

We observe no significant modification of the jet fragmentation in d + 
Au collisions, indicating that cold nuclear matter effects are small or 
absent. We find that hadrons carrying a large fraction of the quark's 
momentum are suppressed in Au$+$Au compared to $p$$+$$p$ and $d$$+$Au. This is 
expected from energy loss of partons in quark gluon plasma. As the 
momentum fraction decreases, the yield of hadrons in Au$+$Au increases, 
eventually showing an excess over the jet fragment yield in $p$$+$$p$ 
collisions. The excess is seen primarily at large angles and is most 
pronounced for hadrons associated to lower momentum direct photons.

To address whether the excess is a result of medium modification of the 
jet fragmentation function or the excess indicates the presence of 
``extra" particles from the medium, we compared to theoretical 
calculations. The calculations suggest that the observed excess arises 
from medium response to the deposited energy. Furthermore, the excess 
particles appear at low $z_T$, corresponding to low associate hadron 
$p_T$. This can be seen in each direct photon $p_T$ bin.



\begin{acknowledgments}

We thank the staff of the Collider-Accelerator and Physics
Departments at Brookhaven National Laboratory and the staff of
the other PHENIX participating institutions for their vital
contributions.  We acknowledge support from the
Office of Nuclear Physics in the
Office of Science of the Department of Energy,
the National Science Foundation,
Abilene Christian University Research Council,
Research Foundation of SUNY, and
Dean of the College of Arts and Sciences, Vanderbilt University
(U.S.A),
Ministry of Education, Culture, Sports, Science, and Technology
and the Japan Society for the Promotion of Science (Japan),
Conselho Nacional de Desenvolvimento Cient\'{\i}fico e
Tecnol{\'o}gico and Funda\c c{\~a}o de Amparo {\`a} Pesquisa do
Estado de S{\~a}o Paulo (Brazil),
Natural Science Foundation of China (People's Republic of China),
Croatian Science Foundation and
Ministry of Science and Education (Croatia),
Ministry of Education, Youth and Sports (Czech Republic),
Centre National de la Recherche Scientifique, Commissariat
{\`a} l'{\'E}nergie Atomique, and Institut National de Physique
Nucl{\'e}aire et de Physique des Particules (France),
Bundesministerium f\"ur Bildung und Forschung, Deutscher Akademischer
Austausch Dienst, and Alexander von Humboldt Stiftung (Germany),
J. Bolyai Research Scholarship, EFOP, the New National Excellence
Program ({\'U}NKP), NKFIH, and OTKA (Hungary),
Department of Atomic Energy and Department of Science and Technology
(India),
Israel Science Foundation (Israel),
Basic Science Research and SRC(CENuM) Programs through NRF
funded by the Ministry of Education and the Ministry of
Science and ICT (Korea).
Physics Department, Lahore University of Management Sciences (Pakistan),
Ministry of Education and Science, Russian Academy of Sciences,
Federal Agency of Atomic Energy (Russia),
VR and Wallenberg Foundation (Sweden),
the U.S. Civilian Research and Development Foundation for the
Independent States of the Former Soviet Union,
the Hungarian American Enterprise Scholarship Fund,
the US-Hungarian Fulbright Foundation,
and the US-Israel Binational Science Foundation.

\end{acknowledgments}



%
 
\end{document}